\documentclass{hobbs}

\usepackage{graphicx,times}
\usepackage[numbers,super]{natbib}
\usepackage{CJK,color}
\usepackage{url}
\usepackage{fancyhdr}

\providecommand{\bm}[1]{\mbox{\boldmath$#1$\unboldmath}}

\begin{document}
\makeatletter
\def\@cite#1#2{\textsuperscript{[{#1\if@tempswa  #2\fi}]}}
\makeatother
\begin{CJK*}{GBK}{song}


   \title{Using the pulsar timing software package, TEMPO2
}


   \volnopage{Vol.9 (2012) No.3, 257--000}
   \setcounter{page}{257}          
   \author{G. Hobbs
   }
   \institute{CSIRO Astronomy and Space Science, Australia Telescope National Facility, P.O. Box 76, Epping,NSW 1710, Australia
   }
   \date{Received~~2012 month day; accepted~~2012~~month day}    
   \abstract{This paper contains details on the algorithms implemented in the TEMPO2 pulsar
timing software package and describes how the software is used. Information is given
on how to download and install the software, use the various interfaces, simulate realistic
data sets and develop the software. The use of TEMPO2 in predictive mode is also
described.
   }


   \lhead{G. Hobbs: ~Using the pulsar timing software package, TEMPO2}

   \maketitle
\end{CJK*}
%
%
\section{Introduction}           
\label{sect:intro}

The \textsc{Tempo2} software package was first released in the year 2006.  It is based on the original \textsc{tempo} software\footnote{see, e.g., \url{http://www.atnf.csiro.au/research/pulsar/tempo}} (hereafter called \textsc{tempo1}) and implements pulsar timing algorithms with a precision and accuracy of $\sim$1\,ns.  \textsc{Tempo2} has already been used for numerous applications including: determining the mass of the Jovian system\cite{cc10}, providing the first evidence for ion-neutral damping in the interstellar medium\cite{y+07}, developing an ensemble pulsar time scale, studying radio pulses from a main-sequence star\cite{r+10}, multi-telescope observations of various pulsars\cite{j08,s+08} and studies of pulsar glitch events\cite{y+10}.

Usually \textsc{tempo2} is used to compare a model for a pulsar's rotation, position and orbital parameters with actual observations of pulse arrival times.  The difference between actual and predicted arrival times are known as the pulsar ``timing residuals''.  After calculating these timing residuals, \textsc{tempo2} carries out a linear least-squares-fit to improve the parameters in the model.   It is possible that the model is too simplistic and does not contain all the phenomena that affect the pulse arrival times.  For instance, the pulsar may undergo glitch events, or may have unmodelled binary companions. Various tools exist  to study such effects that are not included in the timing model.  \textsc{Tempo2} can also be used to predict the pulse period and phase at any given time.  This is commonly used for online systems that need to ``fold'' the incoming data at the predicted topocentric period of the pulsar.

The algorithms are described in brief in the sections below and detailed in the three major \textsc{tempo2} publications: Hobbs, Edwards \& Manchester (2006)\cite{hem06}, Edwards, Hobbs \& Manchester (2006)\cite{ehm06} and Hobbs et al. (2008)\cite{hobbs08}.   Here we provide an example of the \textsc{tempo2} software in use (Section 2), provide an overview of the algorithms (Section 3), details for using and developing the software (Sections 4 and 5) and some important points to note when published results (Section 6).

\section{Example usage of TEMPO2}\label{sec:observations}

In this section we provide an overview of how \textsc{tempo2} is used. For this demonstration we use observations from the Parkes Observatory.   Our procedure for obtaining observations to process is as follows:
\begin{itemize}
\item{Obtain pre-processed data files from the Parkes data archive.  This is described by Khoo (these proceedings) and in Hobbs et al. (2011)\cite{hobbs11}.\endgroup Note: for the following example 34 observations were used.}
\item{Create an analytic standard template from the highest S/N observation:
\begin{verbatim}
% paas -Ci f040405_175214.FTp
\end{verbatim}}
\item{Obtain pulse times of arrival using
\begin{verbatim}
% pat -s paas.std -f tempo2 *.FTp > J1539.tim
\end{verbatim}
}
\end{itemize}

An initial timing model is needed.  This can be obtained from the ATNF Pulsar Catalogue\cite{mhth05} \footnote{\url{http://www.atnf.csiro.au/research/pulsar/psrcat}} by typing in the pulsar's name (J1539$-$5626 for our example) into the ``\verb|Pulsar names|" box and clicking on ``\verb|Get Ephemeris|".   The output ephemeris was copied into a text file called \verb|J1539.par|.

\textsc{Tempo2} can be run by typing:

\begin{verbatim}
% tempo2 -f J1539.par J1539.tim
\end{verbatim}
The output is as follows:
\begin{footnotesize}
\begin{verbatim}
Results for PSR J1539-5626


RMS pre-fit residual = 61071.224 (us), RMS post-fit residual = 61071.224 (us)
Number of points in fit = 34


PARAMETER       Pre-fit                   Post-fit                  Uncertainty   Difference   Fit
---------------------------------------------------------------------------------------------------
RAJ (rad)       4.09817561890493          4.09817561890493          0             0             N
RAJ (hms)       15:39:13.96               15:39:13.96               0             0
DECJ (rad)      -0.985070617217135        -0.985070617217135        0             0             N
DECJ (dms)      -56:26:25.4               -56:26:25.4               0             0
F0 (s^-1)       4.108595092               4.108595092               0             0             N
F1 (s^-2)       -8.1859e-14               -8.1859e-14               0             0             N
PEPOCH (MJD)    48376                     48376                     0             0             N
POSEPOCH (MJD)  48376                     48376                     0             0             N
DMEPOCH (MJD)   48376                     48376                     0             0             N
DM (cm^-3 pc)   175.88                    175.88                    0             0             N
START (MJD)     0                         51021.3077195533          0             51021         N
FINISH (MJD)    0                         54144.03803532            0             54144         N
TZRMJD          0                         52458.4201897289          0             52458         N
TZRFRQ (MHz)    0                         1374                      0             1374          N
TZRSITE         7
TRES            0                         61071.2244943135          0             61071         N
EPHVER          2                         2                         0             0             N
---------------------------------------------------------------------------------------------------

Derived parameters:

P0 (s)      = 0.243392200401334         0
P1          = 4.84930777711516e-15      0
tau_c (Myr) = 0.79578
bs (G)      = 1.0994e+12

Total time span = 3122.728 days = 8.550 years
Finishing off: time taken = 1.82 (s)
\end{verbatim}
\end{footnotesize}

This output provides the pre- and post-fit timing residuals (and, for a weighted fit, the reduced-$\chi^2$ of the fit) and the number of observations included in the fit.    For this example, the pre- and post-fit timing residuals are identical as no fit was carried out.   For each parameter in the timing model, the output contains its name, pre- and post-fit parameter values, the uncertainty of the fitted value and the size of the change in the parameter.  Various derived parameters are listed including the pulse period and its derivative, the pulsar's characteristic age and surface magnetic field strength (all calculated from the pulse frequency and its derivative).  The final text in the output gives the total data span and the time taken for the \textsc{tempo2} software to run.

\begin{figure}
\begin{center}
\includegraphics[angle=-90,width=7cm]{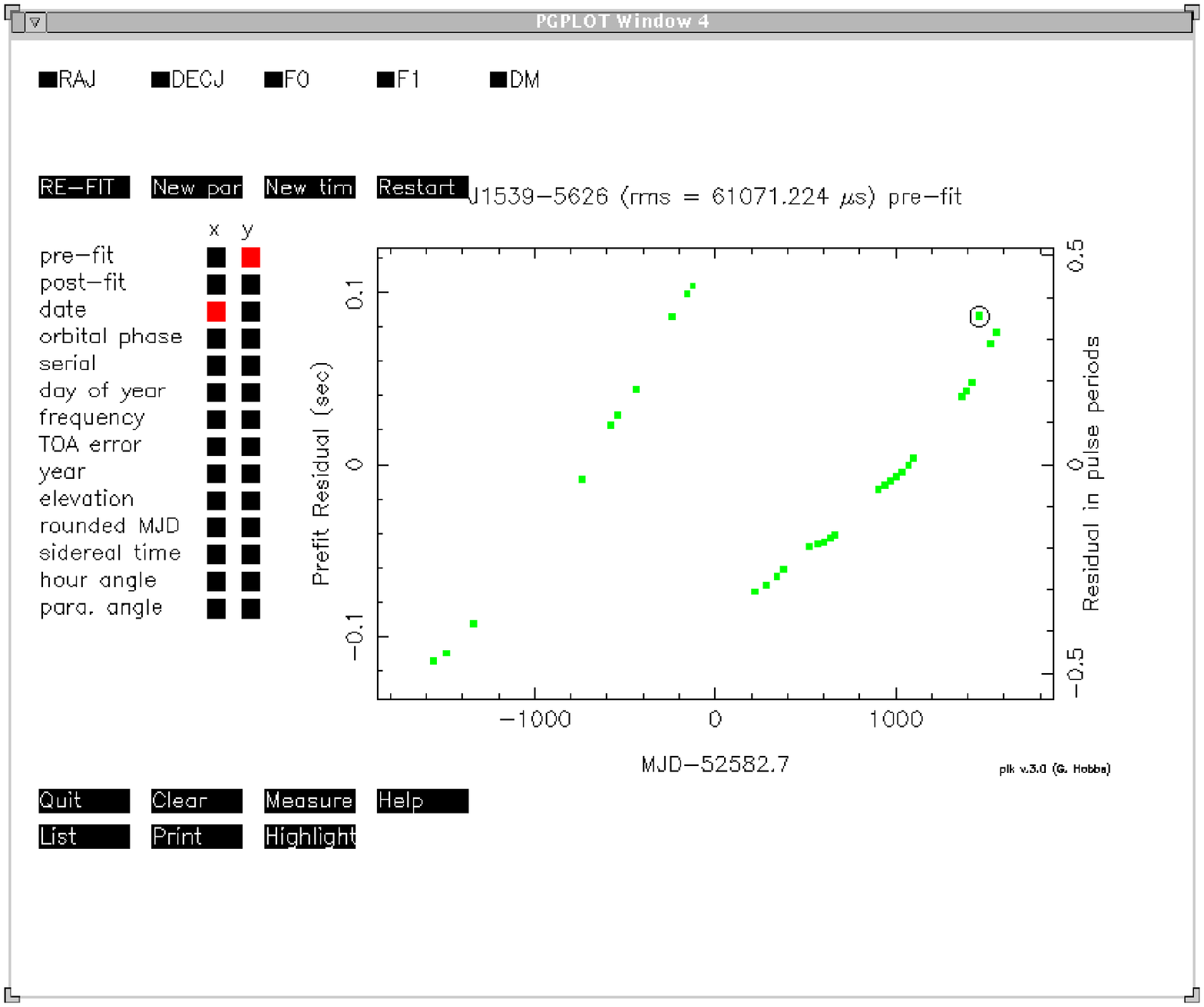}
\includegraphics[angle=-90,width=7cm]{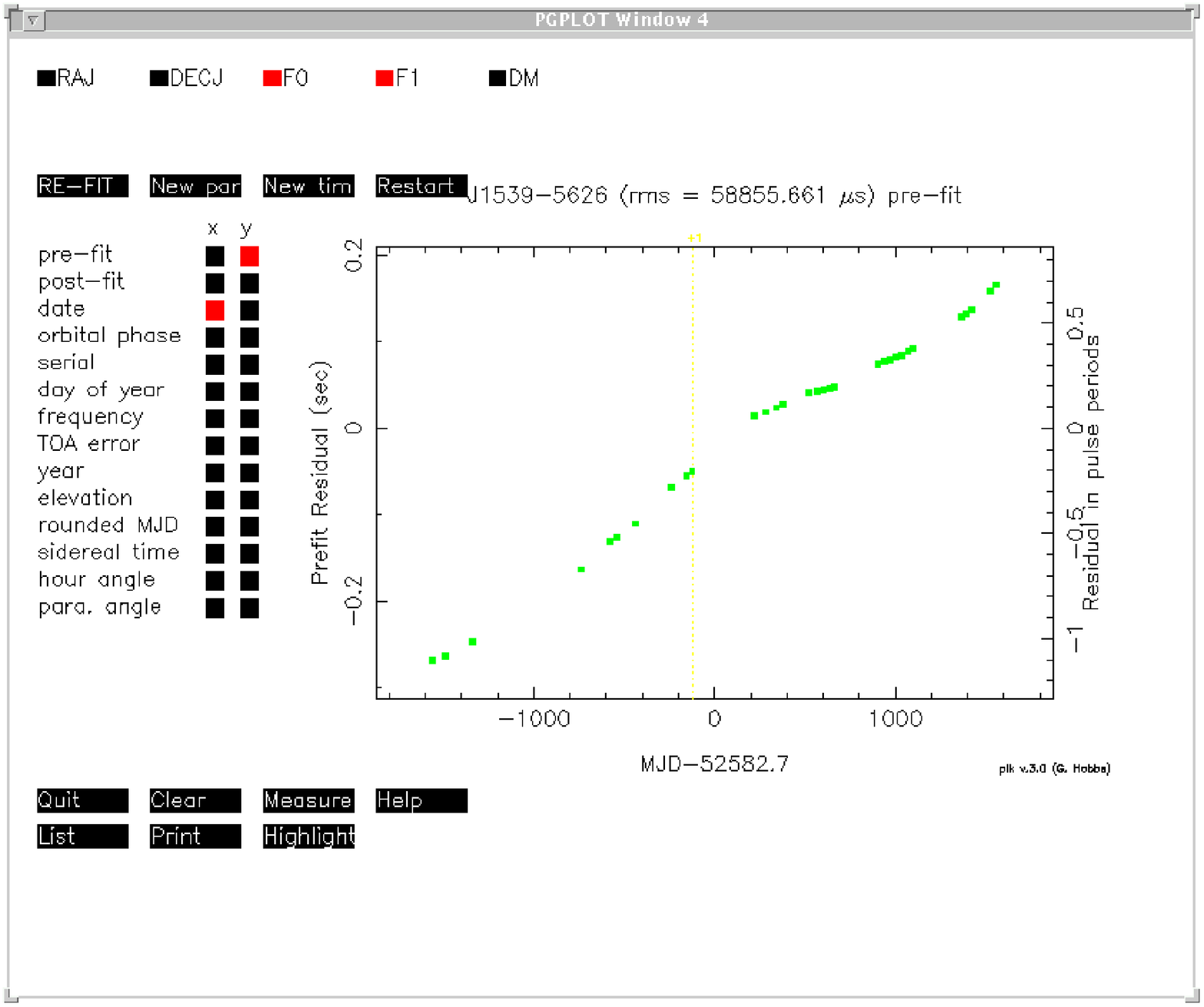}
\includegraphics[angle=-90,width=7cm]{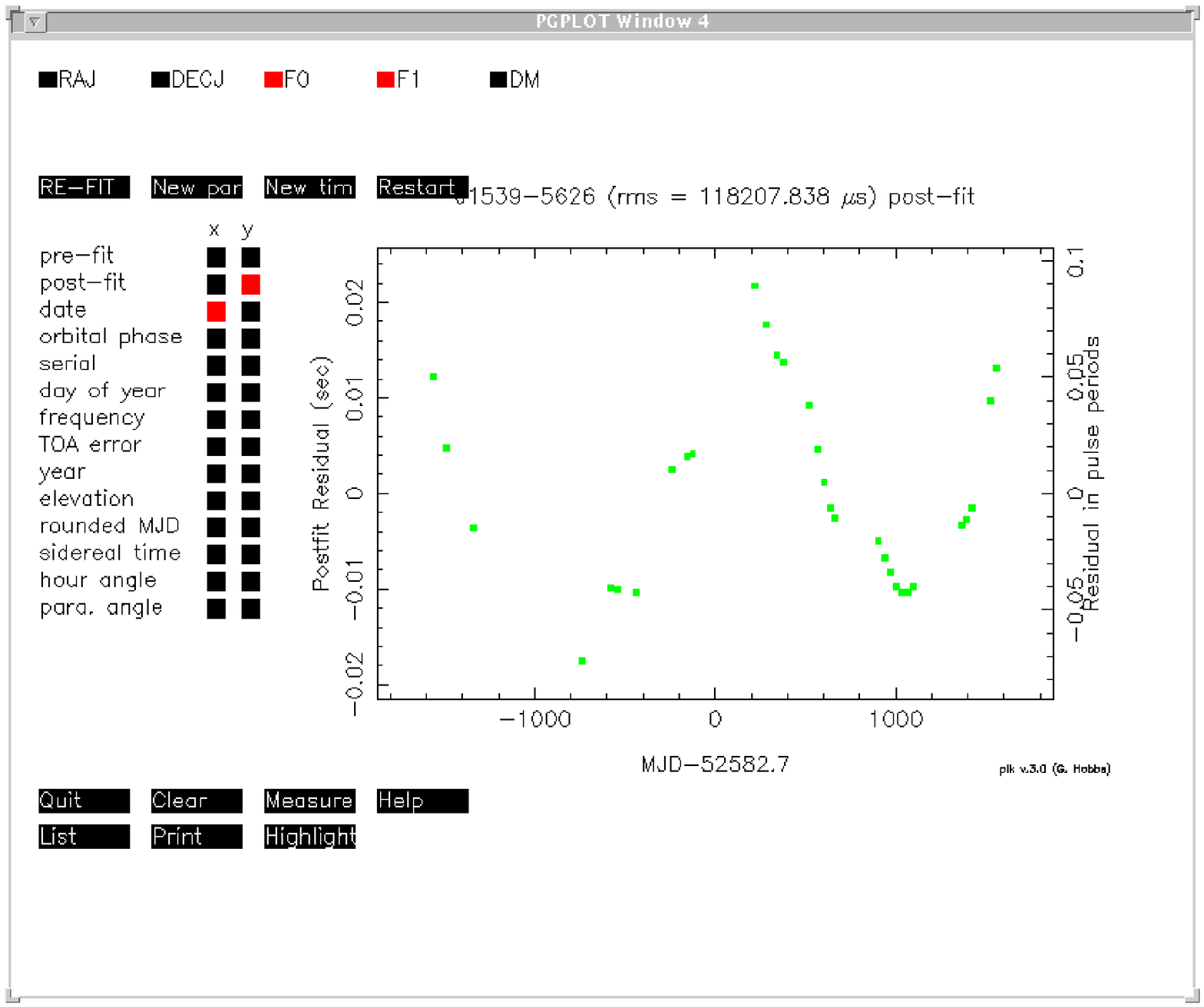}
\includegraphics[angle=-90,width=7cm]{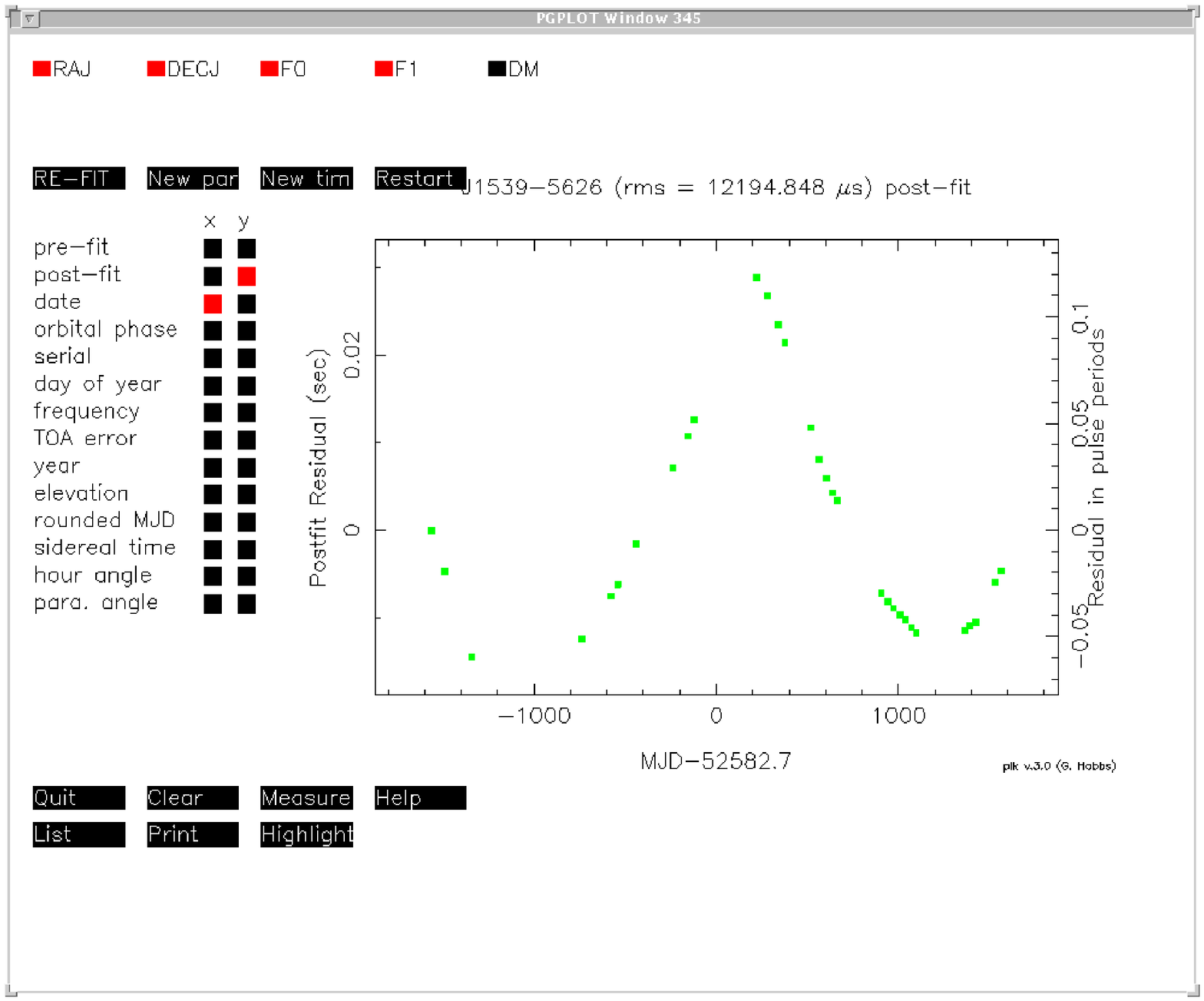}
\end{center}
\caption{Example PSR J1539$-$5626 data set. Each panel in this figure was made using the \textsc{plk} interface to \textsc{tempo2}. The top-left panel contains the timing residuals obtained using the ephemeris in the ATNF pulsar catalogue.  The top-right panel shows the residuals after removing the phase-wrap.  The bottom-left panel shows post-fit residuals after fitting for the pulse frequency and its first time derivative.  The bottom-right panel shows the final fit, which included a fit for the pulsar's position.}\label{fg:1539_ex}
\end{figure}

It is also possible to view the results interactively. Typing

\begin{verbatim}
% tempo2 -gr plk -f J1539.par J1539.tim
\end{verbatim}
produces the output shown in the top-left panel of Figure~\ref{fg:1539_ex} (Note: that white and black has been reversed for these figures).  The small points in the centre of each panel are the ``timing residuals". A group of observations have been circled (by clicking on them with the middle mouse button).  These particular observations are, for an unknown reason, corrupted and should be deleted (by clicking on them with the right mouse button).  Once these points have been removed a new arrival-time file can be created by clicking on ``\verb|New tim|" and typing ``\verb|tempo2|" followed by the new filename (\verb|J1539_2.tim|).

Note from the top-left panel in Figure~\ref{fg:1539_ex} that the residuals have a large offset of one pulse period near the centre of the plot.  This is known as a ``phase wrap" (described in Section~\ref{sec:residuals}).  The phase jump can be removed by positioning the mouse cursor at the time of the phase jump and pressing `+'.   The resulting plot is shown in the top-right panel of the figure.  The main feature of this panel is the linear trend seen in the timing residuals.  This occurs because the pulse frequency in the ephemeris is not precise.   We can update the model for the pulse frequency and its time derivative by selecting the ``F0'' and ``F1'' buttons seen at the top of the page (or, as described above, by placing a `1' in the  timing model file after the parameter value) and then clicking on the ``\verb|RE-FIT|" button.  In the bottom-left panel of the figure, we show the resulting plot after selecting to plot the ``\verb|post-fit|" timing residuals on the y-axis (after fitting, the phase jump can be removed by pressing the ``\verb|Backspace|", or ``\verb|delete|", key).   The post-fit pulsar timing model can be saved by clicking on ``\verb|New par|" and typing in the name of the new parameter file (here we use \verb|J1539_2.par|).

The resulting timing residuals shown in the bottom-left panel of Figure~\ref{fg:1539_ex} are caused by 1) pulsar timing noise\cite{hcmc10} and references therein) and 2) an inaccurate estimate of the pulsar's position in the timing model.  As described in Coles et al. (2011)\cite{chm+12}, it is not correct simply to fit for the pulsar's position in the presence of non-white timing residuals.  Using the methods described in the Coles et al. paper it is possible to determine an improved position using these data.  Putting these values into a new parameter file \verb|J1539_3.par| leads to the timing residuals shown in the bottom-right panel in the figure.

\section{Overview of algorithms}\label{sec:overview}

In this section we provide an overview of the algorithms implemented in \textsc{tempo2}.  Mathematical details are provided in Edwards, Hobbs \& Manchester (2006)\cite{ehm06}.  In brief, the pulse arrival times at the Solar System barycentre are determined. These barycentric arrival times are compared with predicted arrival times given the pulsar timing model to form pulsar timing residuals.  A least-squares-fitting procedure is carried out to improve the pulsar timing model and the entire process iterated to form post-fit timing residuals.

\subsection{Clock corrections and Einstein delay}

Pulse times-of-arrival (TOAs) are measured relative to the observatory time standard (such time standards are usually based on a hydrogen maser providing a stable reference frequency) and the Global Positioning System (GPS) satellites to refer the TOAs to international time standards.   Observatory time standards are usually relatively unstable over timescales of months to years.  It is therefore essential to convert the measured pulse TOAs to a realisation of Terrestrial Time (TT).  For most purposes, International Atomic Time (TAI) is an adequate realisation of terrestrial time, TT(TAI). However, for high precision pulsar timing observations, the post-corrected time standard published by the Bureau International des Poids et Mesures (BIPM) may be necessary.

\textsc{Tempo2} converts the recorded TOAs to a requested realisation of TT via a set of clock corrections.  These clock corrections are given in various text files that contain the difference between two time standards and the date of the measurement.  For instance, to correct observations from the Parkes Observatory, the \verb|pks2gps.clk|, \verb|gps2utc.clk|,  \verb|utc2tai.clk| and \verb|tai2tt_tai| files are used.   For our example data set (used in Section~\ref{sec:observations}), the time corrections from the observatory to terrestrial time range from $\sim 63$ to $65$\,seconds.  This correction arises from an original 32.184\,s offset between TAI and TT, plus leap-second compensation from UTC to TAI (currently 34\,s). The conversion from the Parkes observatory time standard to the GPS standard is typically $\sim$1\,$\mu$s and the conversion from GPS to UTC is $\sim 10$\,ns.

After conversion to a realisation of TT, it is necessary to make the relativistic conversion of the TOAs to the frame of the Solar System barycentre. This is undertaken within \textsc{tempo2} using the tabulated results of Irwin \& Fukushima (1999)\cite{if99}.  The size of this conversion is typically a few milliseconds.

\subsection{Roemer delay}

The Roemer delay, $\Delta_r$, is the geometric time delay between the pulse arriving at the observatory and when the pulse passes the Solar System barycentre.  For a unit vector, $\hat{k}$, pointing from the Earth to the pulsar (assumed to be the same from the barycentre),
\begin{equation}
\Delta_r = \frac{\vec{R} \cdot \hat{k}}{c}
\end{equation}
where $c$ is the vacuum speed of light and $\vec{R}$ is the vector from the observatory to the barycentre.  $R$ is calculated by determining the 1) vector from the observatory to the Earth's centre (obtained from a data file containing the geocentric coordinates of each observatory) and 2) vector from the Earth's centre to the barycentre (obtained from a Solar System planetary ephemeris).   By default \textsc{tempo2} uses the DE405 JPL Solar System planetary ephemeris, but modern precision timing experiments are now making use of DE414 or DE421.

\subsection{Shapiro delay}

The Solar System Shapiro delay\cite{s64} is the time delay caused by the passage of the pulse through gravitational fields in the Solar System.  \textsc{Tempo2} includes the delay caused by the Sun (in which the delay, for a line-of-sight passing the limb of the Sun, is $< 110 \mu$s), Jupiter ($< 180$\,ns), Saturn ($< 58$\,ns), Neptune ($< 12$\,ns) and Uranus ($< 10$\,ns).

For pulsars with a companion, \textsc{tempo2} can also include the effect of the gravitational time delay in the vicinity of the companion (the orbital Shapiro delay); see Section~\ref{sec:orbital}.

\subsection{Dispersive delays}

The pulse TOAs are affected by dispersion in the interstellar medium and in the interplanetary medium.  The interstellar dispersion delay is included in \textsc{tempo2} using the standard relation:
\begin{equation}
\Delta_{\rm ISM} = \frac{D}{(f^{\rm SSB})^2}
\end{equation}
where $D$ is the dispersion constant and $f^{\rm SSB}$ is the frequency of the radiation at the SSB.  The dispersion constant is related to the ``dispersion measure'' (DM) as ${\rm DM} =   2.410\times10^{-16}  D$\,cm$^{-3}$\,pc.  By default, \textsc{tempo2} uses a very simple, spherical model of the Solar Wind to approximate the interplanetary dispersion.  You et al. (2007)\cite{yhcmh07} have implemented a more detailed mode that requires on observations from the Wilcox Solar observatory\footnote{\url{http://wso.stanford.edu/}}.

\subsection{Orbital motion}\label{sec:orbital}

The binary model used by \textsc{tempo2} is based on those of Blandford \& Teukolsky (1976)\cite{bt76}, Damour \& Deruelle (1985)\cite{dd86}, Taylor \& Weisberg (1989)\cite{tw89} with extra terms as described by Kopeikin (1995, 1996)\cite{k95,k96}.  The binary model accounts for the variation in the distance of the pulsar due to the orbital motion, the aberration of the radio beam and the orbital Einstein and Shapiro delays.

For most binary pulsars, the five Keplerian parameters (the binary period, orbital eccentricity, projected semi-major axis of orbit, epoch of periastron and the longitude of periastron) are sufficient to parameterise the orbit.  In some cases, post-Keplerian parameters (such as the periastron advance) are required.   A list of the possible binary parameters is given in Table~7 of Hobbs, Edwards \& Manchester (2006)\cite{hem06}.

\subsection{Forming residuals and the fitting algorithm}\label{sec:residuals}

The delays described in the sections above allow the determination of barycentric arrival times.  The time-evolution of the pulse phase, $\phi$, is determined based on the model of the pulse frequency ($\nu$) and its derivatives:

\begin{equation}
\phi = \phi_0 + \nu t + \frac{1}{2}\dot{\nu}t + \frac{1}{6}\ddot{\nu}t^2 + \ldots
\end{equation}
where $k$ is the number of frequency derivatives and $t$ is the pulse emission time in the reference frame centred on the pulsar.

The timing residual for the i'th observation is determined from:
\begin{equation}
R_i = \frac{\phi(t_i) - N_i}{\nu}
\end{equation}
where $N_i$ is the nearest integer to the value of $\phi$ at the time of the observation, $\phi(t_i)$.  Note that if the pulse arrived earlier than expected, the resulting residual will be negative.

If the model parameter estimates are sufficiently poor that the difference between the predicted arrival time and the actual arrival times becomes greater than half the period of the pulsar then a ``phase wrap'' will occur (as demonstrated in Figure~\ref{fg:1539_ex}).

A linear least-squares-fitting procedure is used to improve the parameter estimates. These new values can subsequently be used to determine post-fit timing residuals.  Usually, this least-squares-fitting procedure accounts for the different arrival time uncertainties between different observations (i.e., it is a weighted fit).  However, if required, it is possible for the user to request an unweighted fit.

\section{Using the software}

\subsection{Downloading and installing the software}

The main \textsc{tempo2} webpage, \url{http://www.atnf.csiro.au/research/pulsar/tempo2}, provides up-to-date details on downloading and installing the software.  \textsc{Tempo2} has been installed on both MacOS X and LINUX operating systems.  It has been tested with recent versions of  \verb|gcc|, \verb|g++| and \verb|gfortran|.  It is also highly recommended that \verb|pgplot| be installed before compiling \textsc{tempo2}.  The most common problems that occur during installation are:
\begin{itemize}
\item{having different versions of \verb|gcc|, \verb|g++| and \verb|gfortran|.  The versions of these packages can be determined using \verb|gcc -v|, \verb|g++ -v| and \verb|gfortran -v|.  Also ensure that your version of \verb|pgplot| has been installed using the same version of \verb|gfortran|.}
\item{not setting the \verb|$TEMPO2| environment variable. Setting of this variable is described in the documentation for installing the software.}
\item{not installing the plugin packages.  It is necessary to type \verb|make install| to ensure that all the plugin packages are copied to the correct directory.}
\end{itemize}

In order to test your installation, type:
\begin{verbatim}
% tempo2 -h
\end{verbatim}
which should provide general information about the software.

\subsection{Required data files}\label{sec:required_files}

\textsc{Tempo2} requires a set of files to run correctly. The most up-to-date versions of these files are part of the standard \textsc{tempo2} distribution, but it may become necessary to update these files.  All the files are stored in a directory structure that is defined by the \verb|$TEMPO2| environment variable.  The most important files are within the following sub-directories:
\begin{itemize}
\item{\verb|clock|: each file has the extension \verb|.clk| and contains the time transfer corrections from one time standard to another.  For instance, the file  \verb|pks2gps.clk| contains the corrections from the Parkes observatory to the global positioning system satellites time standard.  New clock files may be added to this directory as required (for instance, from a new telescope time standard to a realisation of terrestrial time).}
\item{\verb|earth|: This directory contains a single file that contains the Earth orientation parameters as provides by the International Earth Rotation and Reference Systems Service.}
\item{\verb|ephemeris|: Each file contains a planetary ephemeris necessary for determining the position and velocity of the Solar System objects.}
\item{\verb|observatory|: The file \verb|observatories.dat| contains the geocentric coordinates of all the telescopes that can be used in \textsc{tempo2}.  New telescopes may be added as required.}
\end{itemize}

\subsubsection{Parameter files}

\textsc{Tempo2} always requires a model for the pulsar's spin, astrometric, dispersion measure and, for binary systems, orbital parameters.  This model is included in a single file with at least two columns.  The first column contains a label for each parameter and the second column contains the parameter value.  All parameter files are required to contain the pulsar's name (PSRJ), its right ascension (RAJ), declination (DECJ), a pulse-frequency (F0), dispersion measure (DM) and the epoch of the pulse-frequency measurement (PEPOCH)\footnote{It is possible to provide ecliptic coordinates (ELONG, ELAT) instead of RAJ and DECJ, or the pulse period (P0) instead of the frequency (F0).}.  An example is shown below:

\begin{verbatim}
PSRJ          J1539-5626
RAJ           15:39:14.0643716
DECJ          -56:26:26.28676
F0            4.1085950845961150884    1  4e-10
PEPOCH        48376
DM            175.88
\end{verbatim}
The third column for the pulse frequency (\verb|F0|) row is used to request that \textsc{tempo2} fits for this parameter (the `1' in column 3). The fourth column gives the uncertainty on the parameter ($4 \times 10^{-10}$). Neither of these last two columns are required.

Various other parameters are often included such as the pulse-frequency derivative (F1), orbital parameters or definitions of the clock files to use.

\textsc{Tempo2} parameter files are identified with \verb|EPHVER 5| included in the parameter file.   Following the IAU resolution A4 (1991), the \textsc{tempo2} algorithms make use of barycentric coordinate time (TCB).  Pulsar timing models obtained using \textsc{tempo1} made use of barycentric dynamical time (TDB).  The conversion from TDB parameter file to TCB is described in Hobbs, Edwards \& Manchester (2006)\cite{hem06} and can be carried out by typing\footnote{\textsc{Tempo2} can automatically carry out this conversion if the parameter \textsc{EPHVER 2} is included in the parameter file. However, we recommend converting all parameter files to be consistent with TCB.}:

\begin{verbatim}
% tempo2 -gr transform oldfile.par newfile.par
\end{verbatim}
The output file (\verb|newfile.par|) will be compatible with \textsc{tempo2}.

\subsubsection{Arrival time files}

Arrival time files contain the pulse arrival times and related information for each observation.  The start of a typical arrival time file is shown below for Parkes observations.
\begin{verbatim}
FORMAT 1
file1 1405.0   52618.5056325016767 0.236 PKS -r MULTI -b CPSR2
file2 660.89   52742.3494510752533 1.209 PKS -r 50CM -b CPSR2
file3 1431.028 53143.0195601772366 0.109 PKS -r H-OH -b WBCORR
\end{verbatim}

The first line of the file (\verb|FORMAT 1|) indicates that this file is in the standard \textsc{tempo2} format (note, it is possible to read earlier \textsc{tempo1}-style files, if necessary).  Each subsequent line corresponds to a pulse site-arrival-time.  The first five columns are required for all observations and contain the file name or identifier, the observing frequency (MHz), the pulse site-arrival-time (MJD), the uncertainty on the arrival time ($\mu$s) and the observing site.  In the example above, we have also included two ``flags" that represent the receiver used (given by the ``\verb|-r|'' flag) and the observing backend system (the ``\verb|-b|'' flag).  Such flags are not necessary, but can be very useful in subsequent processing to identify observations of interest.  The choice of flag labels (here ``-r'' and ``-b'') can be chosen by the user\footnote{There are a few flag labels, such as ``-dm" that have particular meaning to \textsc{tempo2}.  These are described in the standard \textsc{tempo2} documentation.}.

\subsection{Getting help}

There are various ways to learn how to use the \textsc{tempo2} software.  For most interfaces it is possible simply to type \verb|-h| on the command-line to obtain basic usage information, for instance typing:
\begin{verbatim}
% tempo2 -h
\end{verbatim}
will describe the main \textsc{tempo2} code, whereas:
\begin{verbatim}
% tempo2 -gr plk -h
\end{verbatim}
will describe the \verb|plk| graphical interface (see Section~\ref{sec:plk}).  The main \textsc{tempo2} website provides documentation on many aspects of the software.  However, the webpage also contains tutorials that provide step-by-step instructions on how to do a particular task (such as fitting for pulsar parameters in the presence of pulsar timing noise). Many of the graphical plugin-packages provide interactive help.  This is usually accessed by pressing `h'.

General discussion on the use of \textsc{tempo2} and reports on issues within the software are emailed to the \textsc{tempo2} email distribution list.  This list is publically available from \url{http://pulsarastronomy.net}.

\subsection{The graphical interfaces}

The \textsc{tempo2} functionality can be enhanced using ``plugin-packages'' (or ``interfaces'').  These interfaces can be used to provide different textual output (the ``output'' interfaces; section \ref{sec:output}) or to provide interactive, graphical interfaces\footnote{It is also possible to use interfaces to control the selection of data for analysis (the {\tt -splug} interfaces) and for changing the fitting algorithms within \textsc{tempo2} (the {\tt -fitfunc} interfaces).  These interfaces are not discussed in this paper.}.  In the subsequent sections we describe some of the most commonly-used interfaces.

\subsubsection{plk}\label{sec:plk}

The most common graphical interface is \textsc{plk} which provides an interactive interface for viewing and processing the pre- and post-fit timing residuals.  Figure~\ref{fg:1539_ex} shows a typical example.  By default all observations in the 50\,cm observing band are displayed in red, those at 20\,cm in green and around 10\,cm in blue. The user may select different bands or colours as required (see below).

The plugin provides a large number of options that can be obtained by pressing various keys.  The most common are listed in Table~\ref{tb:plk}.  There are also a set of command-line arguments that may be used:
\begin{verbatim}
% tempo2 -gr plk -f J1539.par J1539.tim -us
\end{verbatim}
will plot the residuals in microseconds (note: \verb|-ns| and \verb|-ms| may also be used to plot in nanoseconds and milliseconds respectively).
\begin{verbatim}
% tempo2 -gr plk -f J1539.par J1539.tim -publish
\end{verbatim}
selects a suitable colour scheme for making publication quality plots.

In order to modify the colours and fonts, a new file can be created that is similar to the  \verb|$TEMPO2/plugin_data/plk_setup_data.dat| file.  The file can be stored in the local directory and is accessed using:

\begin{verbatim}
% tempo2 -gr plk -f J1539.par J1539.tim -setup mysetupfile.dat
\end{verbatim}
The available options for setup files are listed in Table~\ref{tb:setupfiles}.

\begin{table}[h!!!]
\small
\centering
\begin{minipage}[]{120mm}
\caption[]{Commands Available For the ``plk'' Interface}\label{tb:plk}\end{minipage}
\tabcolsep 6mm
 \begin{tabular}{ll}
  \hline\noalign{\smallskip}
Key &  Meaning                \\
\hline
\verb|1| & plot pre-fit residuals versus date \\
\verb|2| & plot post-fit residuals versus date \\
\verb|3| & plot pre-fit residuals versus orbital phase \\
\verb|4| & plot post-fit residuals versus orbital phase \\
\verb|5| & plot pre-fit residuals serially \\
\verb|6| & plot post-fit residuals serially \\
\verb|7| & plot pre-fit residuals versus day of year \\
\verb|8| & plot post-fit residuals versus day of year \\
\verb|9| & plot pre-fit residuals versus frequency \\
\verb|a| & plot post-fit residuals versus frequency \\
\verb|+| & Add phase wrap at mouse position \\
\verb|-| & Subtract phase wrap at mouse position \\
\verb|Backspace| & Remove all phase wraps \\
\verb|g| & Change graphics device \\
\verb|h| & Display `help' information \\
\verb|H| & Highlight observations based on a flag value \\
\verb|l| & List all points in region \\
\verb|o| & Highlight all observations in selection region \\
\verb|N| & Highlight observation based on its filename \\
\verb|w| & Toggle fitting with weights \\
\verb|x| & Re-do the fit \\
\verb|z| & Select a group of observations using the mouse \\
\verb|Z| & Delete a group of observations selected using the mouse \\
\verb|ctrl-d| & delete all selected observations \\
\verb|ctrl-i| & Highlight observations based on their flag values \\
\verb|ctrl-r| & Create a region file for use with the glitch plugin (see Section~\ref{sec:glitch})\\
\verb|ctrl-s| & Overwrite arrival time file \\
\verb|ctrl-w| & Overwrite parameter file with new values \\
middle mouse & Display profile if \textsc{PSRCHIVE}$^a$ installed \\
right mouse & Delete closest observation \\
left mouse & Display information about closest observation \\
  \noalign{\smallskip}\hline
\end{tabular}
~\\
$^a$ http://psrchive.sourceforge.net\cite{hsm04}.
\endgroup
\end{table}

\begin{table}[h!!!]
\small
\centering
\begin{minipage}[]{120mm}
\caption[]{Commands Allowed in a Setup File for the \textsc{plk} Interface}\label{tb:setupfiles}\end{minipage}
\tabcolsep 6mm
 \begin{tabular}{lp{8cm}}
  \hline\noalign{\smallskip}
Key &  Meaning                \\
\hline
\verb|aspect| & Aspect ratio (default = 0.8) \\
\verb|background| & Colour of background (default = ``black'') \\
\verb|freq <flo fhi col style>| & Definition of symbol and colour for a given frequency range.  The first two parameters give the frequency range (MHz).  The remaining two parameters give the PGPLOT colour and point-style. \\
\verb|fontsize| & Size of font (default = 1.0) \\
\verb|fonttype| & Font \textsc{pgplot}-style (default = 1) \\
\verb|line| & Colour of foreground lines (default = ``white'') \\
\verb|linewidth| & Line width (default = 1.0) \\
\verb|menu| & Initial menu display (use 0 for no menu) \\
  \noalign{\smallskip}\hline
\end{tabular}
\end{table}

\subsubsection{glitch}\label{sec:glitch}

The \textsc{glitch} plugin allows glitch events to be analysed and identified.  This plugin requires the input of a set of parameter and arrival time files that provide coherent timing solutions of specified time intervals These can be obtained by creating a ``region'' file using the \textsc{plk} plugin.  Within \textsc{plk} the user first presses 'ctrl-r' and then uses the left mouse button to indicate data spans which include a sufficient number of observations to determine the pulse frequency and, if required, its derivative.  The right mouse button is clicked to complete this process and to create the ``region'' file.

\begin{figure}
\begin{center}
\includegraphics[angle=-90,width=12cm]{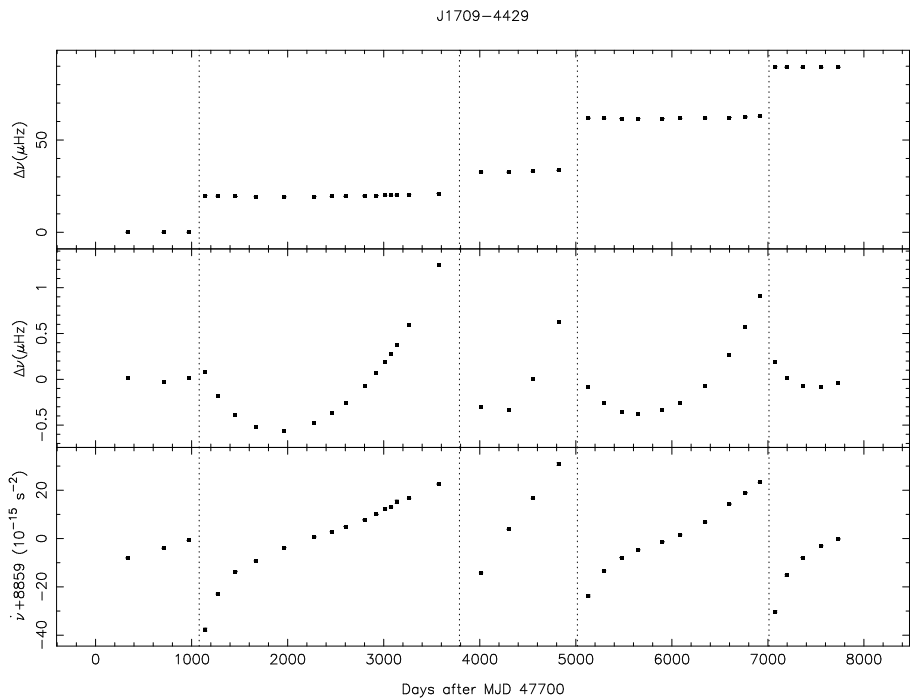}
\caption{Glitch events in PSR~J1709$-$4429. Figure from Yu, M. (personal communication) using the \textsc{glitch} plugin to \textsc{tempo2}.  Each vertical, dotted line indicates the epoch of a glitch. The top panel shows the pulse-frequency as a function of time.  The second panel shows the same, but after a mean has been subtracted between each glitch event.  The bottom panel shows the time variation of the pulse-frequency derivative.}\label{fg:glt}
\end{center}
\end{figure}

For each time interval in the ``region'' file, the glitch plugin determines the pulse frequency and, if requested, its first time derivative.  These values can subsequently be plotted (an example is shown in Figure~\ref{fg:glt}).  Such figures can be used to estimate the time and size of each glitch.  These estimates can subsequently be included in the pulsar's timing model and a standard \textsc{tempo2} fit can be used to obtain improved estimates.

\subsubsection{plotMany and splk}

\begin{figure}
\begin{center}
\includegraphics[width=90mm]{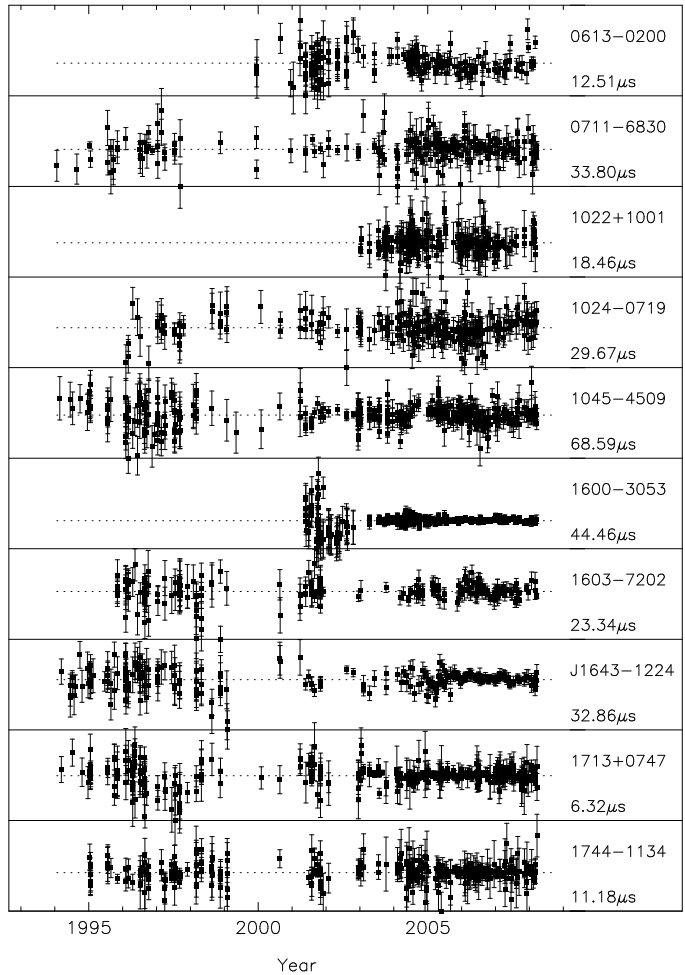}
\caption{Example of the \textsc{plotMany} plugin that shows the post-fit timing residuals for multiple pulsars.  These data were published by Verbiest et al. (2009)\cite{v+09}.\endgroup The value under each pulsar name gives the range from the minimum to the maximum residual.}
\label{fg:plotMany}
\end{center}
\end{figure}

In contrast to \textsc{tempo1}, the \textsc{tempo2} software can process multiple pulsars simultaneously.   Extra pulsar data sets are added to the command line using multiple \verb|-f| arguments.   For instance,
\begin{verbatim}
% tempo2 -gr plotMany -f psr1.par psr1.tim -f psr2.par psr2.tim
         -f psr3.par psr3.tim
\end{verbatim}
will obtain post-fit timing residuals for all three pulsar data sets.  An example of the output from the \verb|plotMany| plugin is shown in Figure~\ref{fg:plotMany}.

If only a few pulsars are being analysed then the \textsc{splk} interface can be used.  This interface allows their pre- and post-fit residuals to be displayed using the same scaling (either in separate panels or overlaid in one figure).

New algorithms are being developed that allow fits to be applied globally over many pulsars.  For instance, Hobbs et al. (2010)\cite{hlk10} reported on determining errors in the terrestrial time standard by fitting for the correlated signal within each pulsar's data set.

\subsubsection{Spectrum}

\begin{figure}
\begin{center}
\includegraphics[angle=-90,width=10cm]{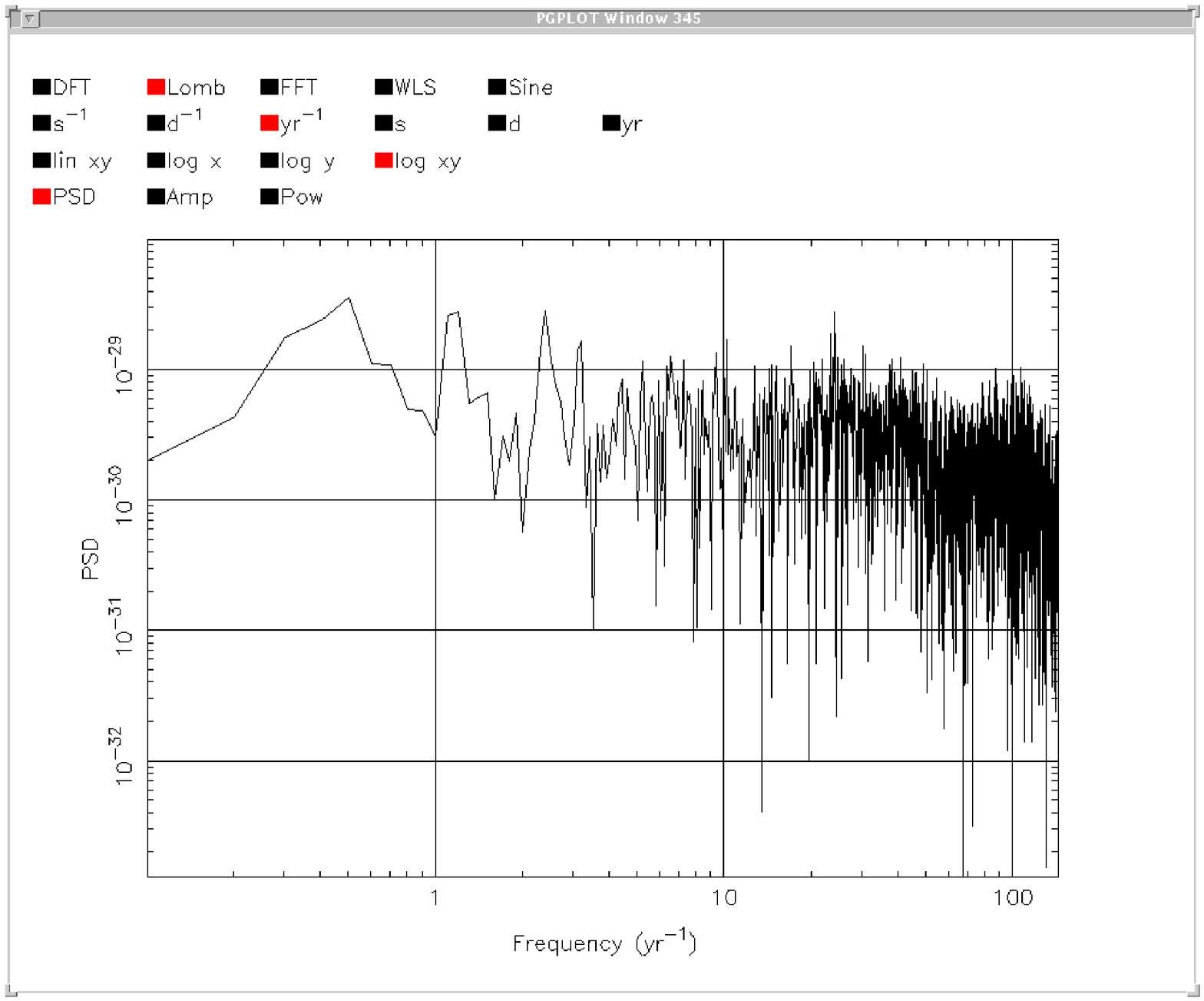}
\caption{Lomb-Scargle power spectrum of the post-fit timing residuals for PSR J0437$-$4715 obtained using the \textsc{spectrum} plugin.}\label{fg:spectrum}
\end{center}
\end{figure}

In order to understand the post-fit timing residuals it is often necessary to obtain a power spectrum of the residuals.  This is not trivial because most pulsar data sets are unevenly sampled.  The \textsc{spectrum} plugin implements various algorithms for determining a power spectrum (including the Lomb-Scargle periodogram and a weighted fit of harmonically related sinusoids).   In Figure~\ref{fg:spectrum} we show the output from:
\begin{verbatim}
% tempo2 -gr spectrum -f 0437.par 0437.tim
\end{verbatim}
where the input data set was taken from Verbiest et al. (2008)\cite{v+08}.

\subsection{Fermi}

The \textsc{Fermi} plugin allows the user to determine a pulse phase for each photon arrival time in a Fermi Large Area Telescope (Fermi) event file.  Documentation and usage instructions for this plugin are available from the Fermi website\footnote{\url{fermi.gsfc.nasa.gov/ssc/data/analysis/user/Fermi_plug_doc.pdf}}.

\subsection{The Output interfaces}\label{sec:output}

For the output interfaces, the \textsc{tempo2} software runs as normal, but, instead of producing the default output, allows the interface to report the results in a manner that is suitable for a particular application.

\subsubsection{General}

The \verb|general| interface provides a generalised method for displaying the post-fit pulsar timing parameters.  This interface can be used to produce output suitable for parsing with other software packages.  For instance,
\begin{verbatim}
% tempo2 -output general -s "{ALL_l} & {ALL_p} \\\\ \n"
      -f mypar.par mytim.tim
\end{verbatim}
will produce output that is suitable for inclusion as a table in a \LaTeX ~document.  If the only parameters of interest are the pulse frequency and dispersion measure then the following can be used:
\begin{verbatim}
% tempo2 -output general -s "Pulse frequency = {F0_v}\nDispersion
       measure = {DM_v}\n" -f mypar.par mytim.tim
\end{verbatim}
A list of the most commonly used commands is given in Table~\ref{tb:general}.

\begin{table}[h!!!]
\small
\centering
\begin{minipage}[]{120mm}
\caption[]{Commands Available For the ``general'' Interface}\label{tb:general}\end{minipage}
\tabcolsep 6mm
 \begin{tabular}{ll}
  \hline\noalign{\smallskip}
Parameter &  Meaning                \\
  \hline\noalign{\smallskip}
\verb|ALL_l|  & List the labels of all parameters in a column  \\
\verb|ALL_v| & List the values of all the parameters in a column \\
\verb|ALL_e| & List the parameter uncertainties in a column \\
\verb|ALL_p| &  Use publication format for displaying parameter uncertainties \\
\verb|F0_p| & List the F0 parameter$^a$\\
\verb|F0_l| & Display the label for the F0 parameter$^a$  \\
\verb|F0_e| & Display the uncertainty for the F0 parameter$^a$ \\
\verb|F0_p| & Display the uncertainty for the F0 parameter in publication mode$^a$ \\
\\
\verb|name| & Pulsar's name \\
\verb|nobs| & Number of observations \\
\verb|postrms| & Rms of the post-fit timing residuals (s) \\
\verb|prerms| & Rms of the pre-fit timing residuals (s) \\
\verb|tspan| & Data span (d) \\

  \noalign{\smallskip}\hline
\end{tabular}
~\\
$^a$ can also be used for any other parameter (e.g., \verb|RAJ_v|, \verb|DM_e|)

\end{table}

\subsubsection{General2}

As described in Section~\ref{sec:overview}, \textsc{tempo2} determines a large number of time delays that need to be applied to the observed site-arrival-times before the timing residuals are calculated.  The \verb|general2| interface provides access to all the delays that are calculated and the various parameters required in the algorithms for each observation.  For instance,
\begin{verbatim}
% tempo2 -output general2 -s "{sat} {bat} {freq} {ism} {roemer}
        {shapiro}\n" -f mypar.par mytim.tim  > outfile.txt
\end{verbatim}
lists, for each observation, the site arrival time, barycentric arrival time, observing frequency, and the delays due to the interstellar medium, the Roemer delay and the Solar System Shapiro delay.  In the above example, the output is stored in a file \verb|outfile.txt| that can subsequently be loaded into other software packages (such as \verb|gnuplot|, \verb|matlab|, \verb|idl|, etc.).   A list of the commonly used parameters that can be used with the \verb|general2| interface are given in Table~\ref{tb:general2}.

\begin{table}[h!!!]
\small
\centering
\begin{minipage}[]{120mm}
\caption[]{Most Common Commands Available with the ``general2'' Interface}\label{tb:general2}\end{minipage}
\tabcolsep 6mm
 \begin{tabular}{llc|}
  \hline\noalign{\smallskip}
Parameter &  Meaning                \\
  \hline\noalign{\smallskip}
\verb|bat|  & barycentric arrival time  \\
\verb|clock0|, \verb|clock1|, \ldots & The various clock corrections (s) \\
\verb|earth_ssb1|  & X-coordinate of distance to Solar System Barycentre from Earth  (lt-s) \\
\verb|earth_ssb2|  & Y-coordinate of distance to Solar System Barycentre from Earth  (lt-s) \\
\verb|earth_ssb3|  & Z-coordinate of distance to Solar System Barycentre from Earth  (lt-s) \\
\verb|err| & Time of arrival uncertainty ($\mu s$)  \\
\verb|file| & identification of pulsar observation (usually filename) \\
\verb|freq| & observing frequency (MHz)  \\
\verb|ipm|  &Delay caused by the interplanetary medium (s)\\
\verb|ism|  & Delay caused by the interstellar medium (s) \\
\verb|pre|  & Pre-fit timing residual (s)  \\
\verb|post|  & Post-fit timing residual (s)  \\
\verb|roemer|  & Roemer delay (s)  \\
\verb|sat|  & site arrival time  \\
\verb|shapiro|  & Shapiro delay caused by Sun (s)  \\
\verb|shapiroJ|  & Shapiro delay caused by Jupiter (s)  \\
\verb|solarangle| & Angle between the Observatory, pulsar and Sun (deg) \\
\verb|tropo|  & Tropospheric delay (s)  \\
\verb|tt|  & Correction from observatory time standard to Terrestrial Time (s)  \\
\verb|tt2tb|  & Correction from Terrestrial Time to Barycentric Time (s)  \\
  \noalign{\smallskip}\hline
\end{tabular}
\end{table}

\subsection{Simulating data}

Data can be simulated for numerous purposes included 1) creating test data sets to trial \textsc{tempo2} or other programs, or 2) producing realistic simulations for use in Monte-Carlo simulations.

The \verb|fake| graphical interface provides an easy method for simulating simple data sets.  For instance,
\begin{verbatim}
% tempo2 -gr fake -f 1534_3.par -ndobs 14 -nobsd 1 -randha y -ha 8
    -rms 1e-3 -start 51000 -end 54000
\end{verbatim}
will simulate regularly sampled data from MJD 51000 to 54000 with one observation (\verb|-nobsd 1|) every 14 days (\verb|-ndobs 14|).  The time of each observation will be based upon the pulsar's transit time.  An assumed hour angle range of 8\,hours (\verb|-ha 8|) is used and a random hour angle chosen (\verb|-randha y|).  Each arrival time has an uncertainty of 1\,$\mu$s (\verb|-rms 1e-3|).  One realisation of such timing residuals is shown in the left-hand panel of Figure~\ref{fg:fake}.  The resulting arrival times are stored in a file called \verb|J1539_3.simulate|.

Note that the \verb|fake| interface can simulate data for numerous pulsars:
\begin{verbatim}
% tempo2 -gr fake -f mypar1.par mypar2.par mypar3.par ...
\end{verbatim}

For non-regularly sampled data, it is possible to provide a list of MJDs (e.g., in a file called \verb|mytimes.dat|):
\begin{verbatim}
% tempo2 -gr fake -times mytimes.dat ...
\end{verbatim}

\begin{figure}
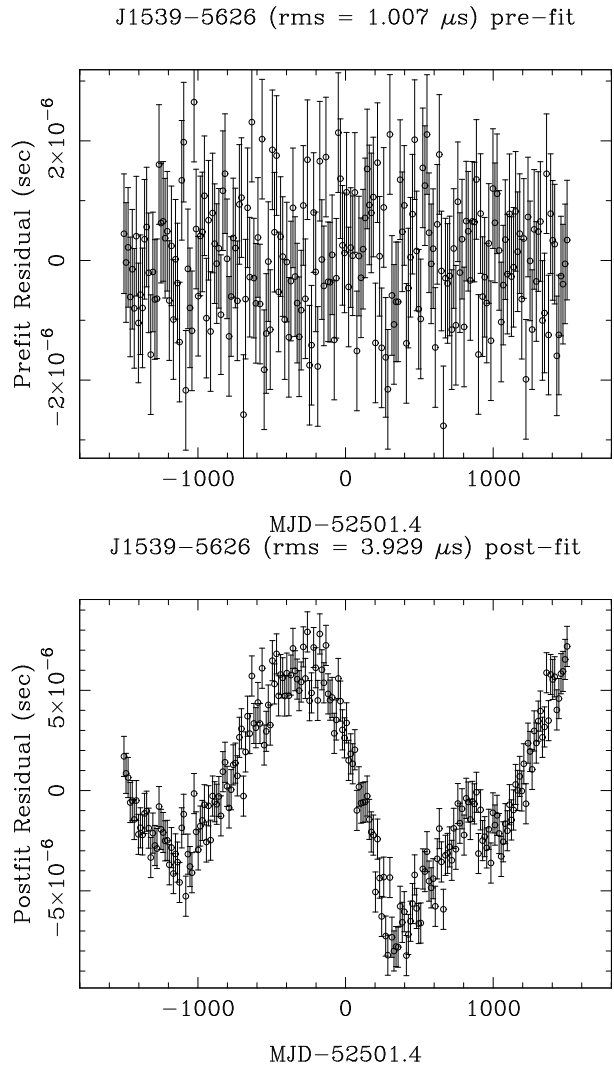

\begin{center}
\includegraphics[angle=-90,width=8cm]{fake_1.ps}
\includegraphics[angle=-90,width=8cm]{fake_2.ps}
\caption{Simulated data sets.  The left panel contains simulated white noise produced using the \textsc{fake} plugin.  The right panel shows the timing residuals induced by a simulated gravitational wave background obtained using the \textsc{GWbkgrd} plugin.}\label{fg:fake}
\end{center}
\end{figure}

\textsc{Tempo2} also provides numerous methods for simulating the induced timing residuals caused by gravitational waves.  The \textsc{GWbkgrd} plugin may be used to simulate the effect of an isotropic, stochastic background of gravitational waves\cite{hcmc10}.   The timing residuals for one realisation of such a background is shown in the right-hand panel of Figure~\ref{fg:fake} and can be obtained using:
\begin{verbatim}
% tempo2 -gr GWbkgrd -f J1539_3.par J1539_3.simulate -alpha -0.666
      -dist 1 -gwamp 1e-13 -ngw 10000
\end{verbatim}
where \verb|-gwamp| gives the dimensionless amplitude of the gravitational wave background, \verb|-alpha| is the spectral exponent of the background, \verb|-dist| the pulsar's distance in kpc and \verb|-ngw| the number of gravitational wave sources to simulate.

\subsection{TEMPO2 and predictive mode}

There are many applications that require knowledge of a pulsar's pulse phase or frequency at a specified time (or over a particular time interval).  For instance, when carrying out observations of pulsars it is common to ``fold'' the incoming data stream at the known pulse frequency.   For such applications both \textsc{tempo1} and \textsc{tempo2} provide a polynomial approximation of the topocentric phase and frequency over specified time intervals.    The number of polynomial coefficients and the time-span to be fitted are defined by the user.

\textsc{Tempo1} produced ``\verb|polycos|'' that are specific to a given observing frequency.  For compatibility with older systems, \textsc{Tempo2} can produce \verb|polycos|, but we instead recommend the use of ``\verb|predictors|''. The two types are similar, but \verb|predictors| are based on Chebyshev polynomials and are both time- and frequency-dependent\footnote{Section 7.1 in Hobbs, Edwards \& Mancheser 2006 provide more details of why a frequency-dependent prediction is necessary for high-precision observations of binary pulsars}.

A predictor can be produced using:
\begin{verbatim}
% tempo2 -f mypar.par -pred  "sitename mjd1 mjd2 freq1 freq2
     ntimecoeff nfreqcoeff seg_length"
\end{verbatim}
(note: use of quote marks around the arguments for the \verb|predictor|).  Here \verb|sitename| is the code for the observatory (see Section~\ref{sec:required_files}), \verb|mjd1| and \verb|mjd2| give the time range for the predictor, \verb|freq1| and \verb|freq2| give the frequency range (in MHz), \verb|ntimecoeff| and \verb|nfreqcoeff| give the number of time and frequency coefficients respectively and \verb|seg_length| gives the duration of each segment (in seconds).  For instance,

\begin{verbatim}
% tempo2 -f J1539_2.par -pred "PKS 53000 53000.1 1400 1420 12 2 3600"
\end{verbatim}
produces a file, \verb|t2pred.dat|, that contains three segments.  The first of which is:
\begin{footnotesize}
\begin{verbatim}
ChebyModelSet 3 segments
ChebyModel BEGIN
PSRNAME J1539-5626
SITENAME PKS
TIME_RANGE 53000 53000.04166666666666785090455960017
FREQ_RANGE 1400 1420
DISPERSION_CONSTANT -2998630.668904615781912070815451443
NCOEFF_TIME 12
NCOEFF_FREQ 2
COEFFS 769018109.217955444648396223783493 3.686485191186269124433103496556908e-10
COEFFS 14791.39113118368792765267016875441 -2.714781880778597648478063579182172e-09
COEFFS -0.0002377901740449791153227876701965922 2.400943518688284986653565320064977e-11
COEFFS 2.0982624846510589122772216796875e-06 1.389117321526047221513094231558003e-10
COEFFS 1.207348153305550416267545931637863e-07 -1.183322162781448724601720341887527e-10
COEFFS 6.039044819772243499755859375e-10 -3.94440720927149548982084277310375e-11
COEFFS -3.443953270713488260820802753443092e-10 -8.917790212265989797173902108842651e-11
COEFFS 8.076312951743602752685546875e-10 -1.972203604635747744500213715836949e-10
COEFFS -3.2014213502407073974609375e-10 -2.195148359942397488845804198694467e-10
COEFFS -4.486840528746445973609011254303865e-11 6.088106779527743036338532078059797e-11
COEFFS 1.085330344115694363932982905389227e-10 1.419129115509635876605659330448103e-10
COEFFS -7.94898369349539279937744140625e-10 -2.246597149628547429848632771661586e-10
ChebyModel END
\end{verbatim}
\end{footnotesize}
Note the 12 rows of coefficients (corresponding to the requested 12 time polynomial coefficients) each with two columns (corresponding to the 2 frequency polynomial coefficients).

The polynomial coefficients, $c_{kl}$, approximate the function $\phi(t,f) - kf^{-2}$ where $\phi$ is the pulse phase at time $t$ and frequency $f$ and $k$ is a constant computed to remove most of the frequency dependence due to interstellar dispersion.   Therefore:
\begin{equation}
\phi(t,f) - kf^{-2} = \sum_{l=1}^{m} T_l(f) \left[\sum_{k=1}^n c_{kl}T_k(t)\right]
\end{equation}
where $T_k$ is the k'th Chebyshev polynomial of the first kind.

All parameter files that are made using \textsc{tempo2} include a set of parameters (\verb|TZRMJD|, \verb|TZRFREQ|, \verb|TZRSITE|) that correspond to an imaginary arrival time (for site \verb|TZRSITE| and observing frequency \verb|TZRFREQ|) that, with the given parameter file, leads to zero residual (i.e., is perfectly predicted by the timing model).  These parameters are used when producing a \verb|predictor| to align the prediction to the timing model.  For pulsars that exhibit a large amount of timing noise the prediction using a simple model for the pulsar may not be sufficient.  In these cases it is common to ``whiten'' the timing residuals by including, for instance, higher order frequency derivative terms in the parameter file.  Such timing models are usually unable to predict the pulse phase and frequency in the future.

\section{Developing the software}

Most users and developers of \textsc{tempo2} will never need to modify the main \textsc{tempo2} source code.  However, it is straightforward to improve the functionality of \textsc{tempo2} by developing new ``plugin interfaces''.   These can be written in the C or the C++ programming languages and, as demonstrated below, simple plugins can be written and installed within minutes.

The next two sections describe how to write an ``output'' interface and a ``graphical" interface.  For an ``output'' interface, the main \textsc{tempo2} software is run as usual (i.e., forming barycentric arrival times, residuals, carrying out the fit and obtaining post-fit timing residuals), but the interface routines are called to display the output.  The ``graphical" interface provides access to all the main \textsc{tempo2} routines and provides a flexible manner in which to develop the software.

\subsection{Writing your own plugin interfaces}

\subsubsection{Writing a new output format}

The text below provides a template for writing a new output format interface:

\begin{verbatim}
#include <stdio.h>
#include <stdlib.h>
#include <string.h>
#include <math.h>
#include <tempo2.h>


extern "C" int tempoOutput(int argc,char *argv[],pulsar *psr,int npsr)
{
  int i;
  printf("Pulse frequency = %.10f\n",
  	(double)psr[0].param[param_f].val[0]);
  printf("Pulse frequency derivative = %.5f\n",
  	(double)psr[0].param[param_f].val[1]);
  printf("Dispersion measure = %.5f\n",
  	(double)psr[0].param[param_dm].val[0]);

  for (i=0;i<psr[0].nobs;i++)
    {
      printf("Site arrival time %d = %g\n",i,
      	(double)psr[0].obsn[i].sat);
      printf("Barycentric arrival time %d = %g\n",i,
      	(double)psr[0].obsn[i].bat);
      printf("Observing frequency %d = %g\n",i,
      	(double)psr[0].obsn[i].freq);
      printf("Post-fit residual %d = %g\n",i,
      	(double)psr[0].obsn[i].residual);
    }
}
\end{verbatim}

After installation (see \S~\ref{sec:install} below) this output plugin (which we assume is called \verb|mycode|) is available by running:
\begin{verbatim}
% tempo2 -output mycode -f mypar.par mytim.tim
\end{verbatim}

The above code shows the most important parameters required (such as obtaining the timing model parameters, number of observations and the residuals).  Note that:

\begin{verbatim}
psr[i].param[param_f].val[j]
\end{verbatim}
contains the j'th time derivative of the frequency parameter (\verb|param_f|) for the i'th pulsar.
\begin{verbatim}
psr[i].obsn[j].residual
\end{verbatim}
contains post-fit timing residual corresponding to the j'th observation of the i'th pulsar.

Table~\ref{tb:pulsar} shows the most important variables available for each pulsar.  The available timing model parameters are listed in Table~\ref{tb:pulsar_params1} and their various quantities in Table~\ref{tb:pulsar_params}.  The variables available for each pulsar observation are given in Table~\ref{tb:pulsar_obs}.

\begin{table}[h!!!]
\small
\centering
\begin{minipage}[]{120mm}
\caption[]{Variables Available for Each Pulsar.}\label{tb:pulsar}\end{minipage}
\tabcolsep 6mm
 \begin{tabular}{lllc|}
  \hline\noalign{\smallskip}
Type & Parameter &  Meaning                \\
  \hline\noalign{\smallskip}
 int &  nobs & Number of observations \\
 int & fitNfree & Number of degrees of freedom in fit \\
 int & nFit       & Number of points in the fit \\
  double & posPulsar[3] & Three-vector pointing at pulsar \\
 double&  chisq & $\chi^2$ value from the fit \\
    \noalign{\smallskip}\hline
\end{tabular}
\end{table}

\begin{table}[h!!!]
\small
\centering
\begin{minipage}[]{120mm}
\caption[]{Common Timing Model Parameters.}\label{tb:pulsar_params1}\end{minipage}
\tabcolsep 6mm
 \begin{tabular}{llc|}
  \hline\noalign{\smallskip}
Parameter &  Meaning                \\
  \hline\noalign{\smallskip}
\verb|param_a1| & Semi-major axis of orbit \\
\verb|param_decj| & Declination \\
\verb|param_dmepoch| & Epoch of dispersion measure measurement \\
\verb|param_ecc| & Orbital eccentricity \\
\verb|param_f| & Pulse frequency \\
\verb|param_finish| & MJD of last observation to be used \\
\verb|param_om| & Longitude of periastron \\
\verb|param_omdot| & Advance of longitude of periastron \\
\verb|param_pb| & Orbital period \\
\verb|param_pbdot| & Orbital period decay \\
\verb|param_pepoch| & Epoch of pulse frequency measurement \\
\verb|param_pmdec| & Epoch of proper motion in declination \\
\verb|param_pmra| & Epoch of proper motion in right ascension \\
\verb|param_posepoch| & Epoch of position measurement \\
\verb|param_px| & Parallax \\
\verb|param_raj| & Right ascension \\
\verb|param_sini| & Sine of orbital inclination angle \\
\verb|param_start| & MJD of first observation to be used\\
\verb|param_t0| & Epoch of periastron \\

     \noalign{\smallskip}\hline
\end{tabular}
\end{table}

\begin{table}[h!!!]
\small
\centering
\begin{minipage}[]{120mm}
\caption[]{Variables Available for Each Pulsar Parameter.}\label{tb:pulsar_params}\end{minipage}
\tabcolsep 6mm
 \begin{tabular}{lllc|}
  \hline\noalign{\smallskip}
Type & Parameter &  Meaning                \\
  \hline\noalign{\smallskip}
 char ** &  label & Full description of parameter \\
 char ** & shortlabel & Shortened label for parameter \\
 long double* & val       & Post-fit value of the parameter \\
 long double* & prefit   & Pre-fit value of the parameter \\
 long double* & err &  Uncertainty on the parameter estimate\\
 int * &  fitFlag &  = 1 if the user has requested fitting for this parameter \\
 int * & paramSet & = 1 if this parameter is included in the timing model \\
    \noalign{\smallskip}\hline
\end{tabular}
\end{table}

\begin{table}[h!!!]
\small
\centering
\begin{minipage}[]{120mm}
\caption[]{Variables Available for Each Pulsar Observation.}\label{tb:pulsar_obs}\end{minipage}
\tabcolsep 6mm
 \begin{tabular}{lllc|}
  \hline\noalign{\smallskip}
Type & Parameter &  Meaning                \\
  \hline\noalign{\smallskip}
 long double &  sat & Site arrival time (MJD) \\
 long double & bat & Infinite frequency Solar System barycentric arrival time (MJD) \\
 long double & bbat  & Arrival time at binary barycentre (MJD) \\
 long double & prefitResidual & Pre-fit timing residual (s) \\
 long double & residual & Post-fit timing residual (s) \\
 double & freq & Frequency of observation (MHz) \\
 double & freqSSB & Frequency of observation in Solar System barycentric frame (Hz) \\
 double & toaErr & Error on site arrival time ($\mu s$) \\
 char * & fname & Observation identifier \\
 char * & telID & Telescope identification code \\
    \noalign{\smallskip}\hline
\end{tabular}
\end{table}

\subsubsection{Writing a new graphical interface}

Whereas an ``output'' interface allows the user to change the output display, the graphical interfaces allow the user to control all the fitting and processing that \textsc{tempo2} carries out.  A template for such an interface is given below:
\begin{verbatim}
#include <stdio.h>
#include <string.h>
#include <stdlib.h>
#include <math.h>
#include <tempo2.h>

using namespace std;

extern "C" int graphicalInterface(int argc,char *argv[],
          pulsar *psr,int *npsr)
{
  char parFile[MAX_PSR][MAX_FILELEN];
  char timFile[MAX_PSR][MAX_FILELEN];
  int i;
  double globalParameter;

  *npsr = 1;  /* For a graphical interface
  	          that only shows results for one pulsar */
  printf("Graphical Interface: name\n");
  printf("Author:              author\n");
  printf("CVS Version:         $Revision $\n");
  printf(" --- type 'h' for help information\n");

  /* Obtain all parameters from the command line */
  for (i=2;i<argc;i++)
    {
      if (strcmp(argv[i],"-f")==0)
        {
          strcpy(parFile[0],argv[++i]);
          strcpy(timFile[0],argv[++i]);
        }
    }

  /* Load the parameters       */
  readParfile(psr,parFile,timFile,*npsr);
  /* Load the arrival times    */
  readTimfile(psr,timFile,*npsr);
  preProcess(psr,*npsr,argc,argv);

/* Do two iterations for pre- and post-fit residuals*/
  for (i=0;i<2;i++)
  {
      /* Form the barycentric arrival times */
      formBatsAll(psr,*npsr);
      /* Form the residuals                 */
      formResiduals(psr,*npsr,1);
       /* Do the fitting     */
      if (i==0) doFit(psr,*npsr,0);
      /* Display the output */
      else textOutput(psr,*npsr,globalParameter,0,0,0,"");
   }
// Can add your code here to display the results
  return 0;
}


\end{verbatim}

\subsubsection{Installing your plugin}\label{sec:install}

The plugin can be installed simply with a command such as:
\begin{verbatim}
$ g++ myplug_plug.C -shared -o myplug_${LOGIN_ARCH}_plug.t2
$ cp myplug_${LOGIN_ARCH}_plug.t2 $TEMPO2/plugins
\end{verbatim}
Note that the environment variable \verb|${LOGIN_ARCH}| must be set to the system architecture.

\section{Publishing results from \textsc{Tempo2}}

When publishing new pulsar parameters obtained with \textsc{tempo2} it is essential to describe 1) whether  a weighted or unweighted fit was carried out, 2) the Solar system ephemeris used and 3) details of the clock correction process.  To ensure that the fit has converged, the fitting process should have been iterated until the pre- and post-fit values are identical.

It is not common for the main algorithms within \textsc{tempo2} to change, but it is possible that errors will be found and corrected. Of course, it is also likely that \textsc{tempo2} will be developed and enhanced in the future. In order to ensure that your published results are reproducible various command line arguments can be used:
\begin{verbatim}
$ tempo2 ... -reminder
\end{verbatim}
This appends to a file in the local directory (\verb|T2command.dat|) a description of the exact set of commands used when running \textsc{tempo2}.
\begin{verbatim}
$ tempo2 ... -allInfo
\end{verbatim}
This provides information on exactly which clock correction files are being applied in the conversion to terrestrial time.
\begin{verbatim}
$ tempo2 ... -displayVersion
\end{verbatim}
This displays the version number of every algorithm being used within \textsc{tempo2}.

\section{Conclusions}
\label{sect:conclusion}

\textsc{Tempo2} provides a versatile means to use (and develop) pulsar timing algorithms.   \textsc{Tempo2} will continue to be developed into the foreseeable future.  It is likely that the main \textsc{tempo2} algorithms will not significantly change until nanosecond timing precision is reached with e.g., the Square Kilometre Array or similar future telescopes.  However, it is expected that a large number of plugin interfaces are developed to enhance the capabilities of \textsc{tempo2}.  In the near future it is likely that such interfaces will provide the ability to, e.g., correct for dispersion measure variations, produce a pulsar time-scale and to allow spacecraft to navigate using pulsar observations.

\normalem
\begin{acknowledgements}
This work has been carried out as part of the Parkes Pulsar Timing Array project.   The Parkes radio telescope is part of the Australia Telescope which is funded by the Commonwealth of Australia for operation as a National Facility managed by CSIRO.   GH is the recipient of an Australian Research Council QEII Fellowship (project \#DP0878388) and acknowledges support from the Chinese Academy of Sciences \#CAS KJCX2-YW-T09 and NSFC 10803006.   The \textsc{tempo2} software has been developed by numerous people. In particular, GH acknowledges the work by R. Edwards, R. Manchester, W. Coles, X. You, M. Keith, F. Jenet, D. Yardley, A. Archibald and J. Verbiest.  We thank Han Jun for testing the procedures described in this paper.

\end{acknowledgements}

\label{lastpage}


\end{document}